Title: INFERRING DISEASE CORRELATION FROM HEALTHCARE DATA

CORRESPONDING AUTHOR: Ashish Anand, Department of Computer Science & Engineering, Indian Institute of Technology Guwahati, Assam 781039, India; Email- anand.ashish@iitg.ernet.in; Phone- +91 – 361 258 2374

CO-AUTHOR: Gargi Priyadarshini, Department of Computer Science & Engineering, Indian Institute of Technology Guwahati, Assam 781039, India; Email – gargi.tezpur@gmail.com

**ABSTRACT:**

**Objectives:**

Electronic Health Records maintained in health care settings are a potential source of substantial clinical knowledge. The massive volume of data, unstructured nature of records and obligatory requirement of domain acquaintance together pose a challenge in knowledge extraction from it. The aim of this study is to overcome this challenge with a methodical analysis, abstraction and summarization of such data. This is an attempt to explain clinical observations through bio-medical and genomic data.

**Materials and Methods:**

Discharge summaries of obesity patients were processed to extract coherent patterns. This was supported by Machine Learning and Natural Language Processing based technologies and concept mapping tool along with biomedical, clinical and genomic knowledge bases. Semantic relations between diseases were extracted and filtered through Chi square test to remove spurious relations. The remaining relations were validated against biomedical literature and gene interaction networks.

**Results and Discussion:**

A collection of binary relations of diseases was derived from the data. One set implied co-morbidity while the other set contained diseases which are risk factors of others. Validation against bio-medical literature increased the prospect of correlation between diseases. Gene interaction network revealed that the diseases are related and their corresponding genes are in close proximity.

**Conclusion:**

This study focuses on deducing meaningful relations between diseases from discharge summaries. For analytical purpose, the scope has been limited to a few common, well-researched diseases. It can be extended to incorporate relatively unknown, complex diseases and discover new traits to help in clinical assessments.

# Introduction:

Modern health care industry has been revolutionized with automated systems and large volume of digitized data, resulting in widespread usage of Electronic Health Records (EHRs). An EHR contains all the information accumulated during the course of observation and treatment of patients in health care settings [1]. Healthcare or clinical data exists in various formats, ranging from structured text and narratives to bedside monitor trends; the scope of this study is however confined to text data alone. Clinical text includes narratives written by clinicians like discharge summaries, lab reports, inpatient records, outpatient records etc. Analyzing such data will elucidate coherent patterns which can assist in clinical decision support. Significant clinical knowledge uncovered from such records can also contribute towards building structured knowledge base. However, the inherent traits of this massive volume of data impose challenges as discussed below:

- Semi-structured documents without any generic format and with content which may not be well-formed and grammatically correct hinders in using conventional methods of text parsing.
- Amalgamation of medical terms with non-medical terms to a great extent requires proper demarcation of concepts while extracting information from these.
- The existing knowledge bases and vocabularies for clinical concepts differ in their individual terminologies. There is no normalization for such cases till date.
- Analysis through automated systems require suitable corpus annotated by domain experts, which is very rare in this field.

Machine Learning (ML) has been instrumental in analysis of clinical data ever since automated systems have been adopted for rendering better healthcare. The significance of Natural Language Processing (NLP) techniques in clinical data analysis domain has been extensively reviewed in [2] through an illustration of some of the earliest attempts in designing Clinical Decision Support systems. Some of the endeavors discussed here include an Automatic Reminder Systems which processes narratives in patient charts, a diagnostic system providing diagnostic decision support along with event alerts and a Computerized Provider Order Entry system which takes structured input. Each of these systems had ample scope for improvement through extensive use of NLP. The last few years have witnessed ML being channelized largely to extract hidden knowledge from clinical text. State-of-the-art methods like Conditional Random Fields (CRF) and Support Vector Machine (SVM) have been extensively used by the authors of[3] to identify medication events and connect the related parameters like dosage, frequency etc. of every event. CRF and SVM have also been used in[4] to identify concept boundaries and concept type of clinical entities. To extract clinical concepts and semantic relations between them, the authors of[5] took a semi-supervised approach through Maximum Entropy Modelling, on a set of contextual and syntactic features. In[6] also, similar features have been used along with semantic vectors for relation extraction. The authors of [7] have endeavored to identify inter-sentential relations between the entities found in an EHR, with the use of SVM.

Although Clinical Data Analysis have scaled great heights through such state-of-the-art approaches, there has been less emphasis on correlating and deducing knowledge from large collection of records, in contrast to mere information extraction from an individual record. One of the latest endeavors towards pattern recognition in longitudinal EHRs is the system designed in[8]. This system mines common events in historical EHRs and correlates them with patient outcome. Nevertheless, most approaches designed till date focus only on extracting clinical information from data, whereas such information will be substantial only when the acquired knowledge can be used for clinical decision support in new instances. The objective of this work is to venture into the relatively unexplored domain of deducing the correlations between information collected from various instances and usage of the knowledge already acquired to handle new instances appropriately. The idea is to deduce correlations like, inception of a specific disease *'a'* can lead to some other disorder *'b'*, so that in future instances of a patient being inflicted by disease *'a'* can automatically direct clinicians to adopt necessary steps for disease *'b'* also.

Similar relations like a specific set of symptoms 'x' indicate the presence of a disease 'a' or medicine 'm' is preferable for symptoms 'x' etc. can be inferred. The objective of deriving such correlations lies in uncovering clinical interconnections automatically from discharge summaries. Further, this study is an attempt to explore the bio-medical perspective of the derived relations through archived literature and gain insight into the clinical correspondences between diseases at the molecular level of interactions. The intention behind this entire effort is to obtain clinical relations which can be of substantial use in directing the process of clinical decision making. This could in turn ensure better health care in the long run. However, the clinical entities and corresponding relations in the scope of this work are confined to diseases and correlation between diseases alone, keeping the scope to extend it further to cover other clinical entities and relations.

**MATERIALS AND METHODS:**

Clinical narratives were administered through language processing techniques and concept mapping tools. The deduced relations were filtered through a multi-step validation process.

**Data Source**:

The Research Patient Data Repository at Partners HealthCare under i2b2 framework [9] was used as data source for EHRs. The obesity challenge data available in this repository contains discharge summaries of patients suffering from obesity and other related cardiovascular disorders. In view of the guidelines of Health Insurance Portability and Accountability Act (HIPAA), which states that clinical data should be freed from any sensitive and private information related to patients before any other analysis on it [10], these records had been de-identified. The text in these records is anonymized to mask all sensitive information, before making these accessible to public. For factual information on clinical entities, the UMLS Metathesaurus and Semantic Network of the UMLS framework [11] were used.

Biomedical literature to validate the results was obtained through PubMed [12]. Genomic information on diseases was obtained from OMIM [13] and DisGeNET [14] . BioGRID [15] interaction network of human genes was analyzed through for genetic interactions. For text processing and concept identification, NLTK [16] and MetaMap [17] tools were used.

**Design:**

The records obtained from the repository being already de-identified, no text processing was done explicitly for de-identification and the text was used as available. The de-identified records were analyzed to extract diseases and correlations between diseases. The derived relations were then subjected to a multi-step validation process. At every step, relations get discarded and eventually a few relations which satisfied all the validation criteria were retained as meaningful clinical rules. As an illustration, an excerpt from a discharge summary is shown in Figure 1. Diseases like *Anemia* and *Diabetes Mellitus* will be identified and inferred that these two diseases can co-occur, through the method illustrated in Figure 2.

> PRINCIPAL DIAGNOSIS: Coronary Heart Disease, Anemia and GI bleed.
>
> SECONDARY DIAGNOSES: Diabetes Mellitus type II, chronic kidney disease, Hypertension.
>
> HISTORY OF PRESENT ILLNESS: The patient is an 86-year-old woman with a history of diabetes , chronic kidney disease , congestive heart failure with ejection fraction of 45% to 50% who presents from clinic with a chief complaint of fatigue and weakness for one week. She ruled out for myocardial infarction.
>
> PAST MEDICAL HISTORY: kidney disease ,diabetes mellitus, CAD, atrial fibrillation, status post right total hip replacement approximately 13 years ago
>
> FAMILY HISTORY: Family history of coronary disorder and diabetes mellitus.
>
> She was discharged with plan for regular follow up with her outpatient cardiologist.

Figure 1: Excerpt from a Discharge Summary

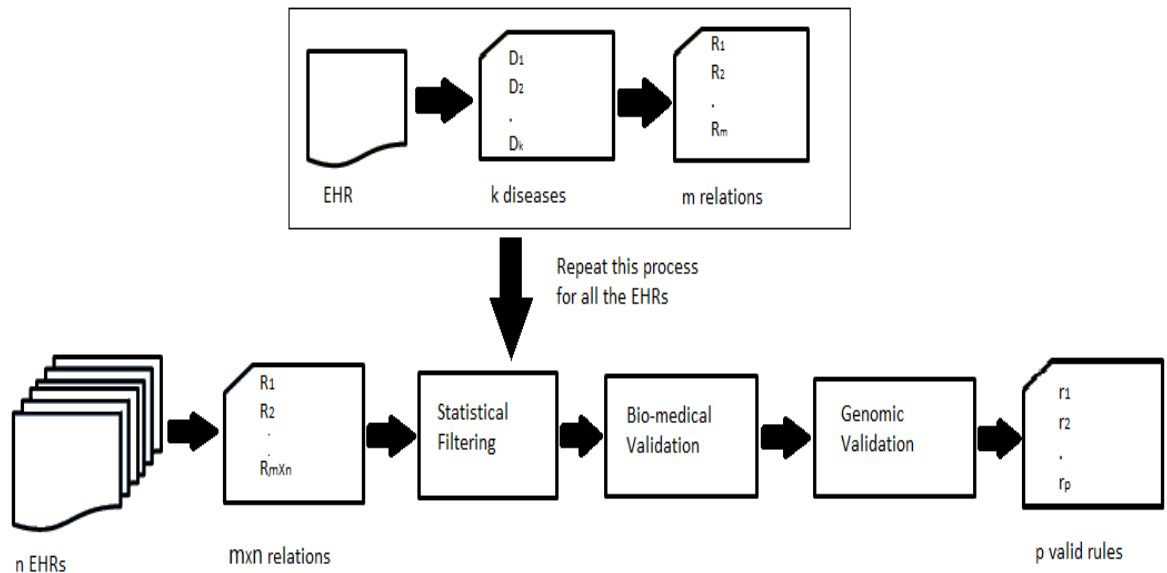

Figure 2: Sequence of operations in the entire method

**Methods:**

Step1 - Extracting Diseases:

Diseases in a discharge summary were extracted with the MetaMap tool followed by two levels of normalization on its output. MetaMap attempts to map every term encountered in the document with some standard clinical concept in the UMLS. In the EHR shown in Figure 1, the mapping generated for the term "CAD" is explained below:

*1000   C0010068:  CAD (Coronary Heart Disease)   [Disease or Syndrome]*

- CAD: common name which is the term as mentioned in the summary
- Coronary Heart Disease: preferred name as in standard UMLS terminology.
- Disease or Syndrome: semantic type of the UMLS concept

The same disease having a standard *preferred name* is referred by various *common names* in different records and also at multiple places of the same record. In the first level of normalization, a mapping was maintained for all *common names* corresponding to a specific *preferred name*. MetaMap often maps the same disease to different concepts of UMLS i.e. different *preferred names*. So in the second level of normalization, all such *preferred names* representing the same concept were replaced with one distinct *preferred name*. Additional processing was done to remove all the negated concepts from the list of diseases. Information available in UMLS on obesity was the only external knowledge source used to support the normalization procedure.

Step 2 - Assigning Labels to Diseases:

The diseases extracted from a record need to be labeled to indicate their context of occurrence. In case of availability of data where diseases are labeled by domain experts, the same can be used to train any competent supervised learning technique so that the predictor generated could be used for all future instances. However, absence of such annotated data lead to the application of a rule based approach for this task. The global context of occurrence of a disease in a discharge summary was captured through its Section Headings. A Section Heading summarizes the content in its respective section and occurrence of a disease under a specific section heading indicates its context of manifestation in the patient. Following features were used for every disease to indicate its presence/absence under specific section headings:

- *Present Medical Status* to indicate presence/absence in sections for current health status.
- *Past Medical Status* to indicate indicating status in all sections concerning the disorders suffered in the past.
- *Diagnosis Results* implying presence/absence in any section describing physical examination and lab tests, diagnosis, medication and treatment.
- *Family History* signifying its mention in sections describing the family history.

Based on the values of these features, every disease was labeled as *Present, Found, Relapse, Family, Inherited, Past, Continuing* or *Unknown.*

Step 3 - Identifying Relations:

With the labeled diseases, binary relations were built for every pair of diseases in a discharge summary. From every pair, a tuple of four elements was created; the first two elements being the two diseases and the next two being their individual labels. Depending upon the two labels, a semantic relation was inferred between the concerned diseases, and finally a tuple of three elements consisting of the two diseases and their semantic relation was obtained. For example, if a patient was inflicted by Hypertension in the past and has now suffered from Heart Failure, the initial tuple will be (Hypertension, Heart Failure, Past, Present). Relation will be assigned as (Hypertension, Heart Failure, Risk factor). This work focuses on extracting two types of semantic relations:

- **Co-morbidity** where both the diseases occur together in the same patient. This included all those tuples where labels of both the entities indicated their concurrent existence in the patient.
- **Risk factor** where one disease is a plausible cause for the inception of the other. This included all such instances where the patient had suffered or has been suffering since long time from one disease and the other one has manifested recently.

Eventually a collection of co-morbid diseases and risk factors of diseases were obtained from the data which was validated through the methods described subsequently.

Step 4 - Statistical Processing**:**

A collection of 1,118 EHRs were processed to obtain 9818 co-morbid pairs and 8190 causal pairs. To filter out the spurious pairs from this collection, Chi Square Statistics [18] was used. This was computed using the following 2 X 2 Contingency Tables:

| (D1 present) AND (D2 present) | (D1 present) AND NOT (D2 present) |
| --- | --- |
| NOT (D1 present) AND (D2 present) | NOT (D1 present) AND NOT (D2 present) |

Table 1: Co-morbid pairs (D1 and D2 are co-morbid)

| (C in past) AND (E now) | (C in past) AND NOT (E now) |
| --- | --- |
| NOT (C in past) AND (E now) | NOT (C in past) AND NOT (E now) |

Table 2: Risk Factor (C is risk factor and E is effect)

All pairs having p-value < 0.001 were considered as valid. However, number of such relations being large (75 pairs), only a few having very high Chi Square value were chosen for further validation.

| Semantic Relation | Disease 1 | Disease 2 | Chi-Square value |
| --- | --- | --- | --- |
| Co-morbid pair | Hypertension | Diabetes Mellitus | 157.313 |
| | Asthma | Obstructive Sleep Apnea | 110.342 |
| | Diabetes Mellitus | Hypercholesterolemia | 82.555 |
| | Obesity | Hypertension | 25.97 |
| Causal pair | Hypertension | Kidney Diseases | 27.534 |
| | Anemia | Left Ventricular Hypertrophy | 23.445 |
| | Hypertension | Heart Failure | 13.295 |
| | Obesity | Coronary Heart Disease | 9.824 |

Table 3: Chi Square values of few of the pairs of diseases

Step 5 - Bio-medical Validation:

The pairs retained after statistical processing were filtered through UMLS Metathesaurus look-up. Existence of relations like "clinically associated with", "co-occurs with" etc. between two diseases in UMLS increased the probability of the diseases being related and corresponding pairs were retained. The selected relations were validated against facts archived in bio-medical literature retrieved through PubMed. To obtain relevant conference and journal articles about the concerned diseases, queries were designed according to the following pattern:

*< Disease 1 > [MAJR] AND < Disease 2 > [MAJR] AND < relation keyword >*

From the abstracts of the entire set of retrieved articles, sentences indicating meaningful relations were extracted through pattern matching expressions like the following (<w> indicates any arbitrary set of words):

- *<w> Disease 1 <w> Disease 2 <w> co-occur <w>*

- *<w> <Disease 1 patients> <w> <exhibiting> <w> <Disease 2> <w>*

- *<Disease 1> <w> <crucial risk factor> <w> < Disease 2> <w>*

The total number of articles and sentences obtained varied for every pair of diseases. So in the entire collection of sentences, the proportion of sentences aligning with the patterns was computed and then relations were filtered on the basis of this percentage.

| Semantic Relation | Disease 1 | Disease 2 | Sentences |
|---|---|---|---|
| Co-morbid pair | Diabetes Mellitus | Hypercholesterolemia | 74.5% |
| | Asthma | Obstructive Sleep Apnea | 67.5% |
| | Hypertension | Diabetes Mellitus | 69.8% |
| | Obesity | Hypertension | 63.53% |
| Causal pair | Hypertension | Kidney Diseases | 70.7% |
| | Hypertensive disease | Heart Failure | 66.4% |
| | Obesity | Coronary Heart Disease | 83.9% |
| | Anemia | Left Ventricular Hypertrophy | 70.7% |

Table 4: Percentage of sentences indicating relation between diseases

Step 6 - Genomic Validation:
The relations retained after bio-medical validation were further analyzed at the molecular level through the gene interaction network. Genes in close proximity are likely to influence each other in various metabolisms. Consequently, the diseases associated with these genes share some clinically significant relations. So, the BioGRID network was analyzed for such genetic interactions between the genes of the diseases participating in the clinical relations. Some of the results obtained are illustrated in Figure 3 and Figure 4. Each of these depicts a small fragment obtained from the massive network of genes where genes associated with two specific diseases are seen to be the immediate or one hop neighbors of each other.

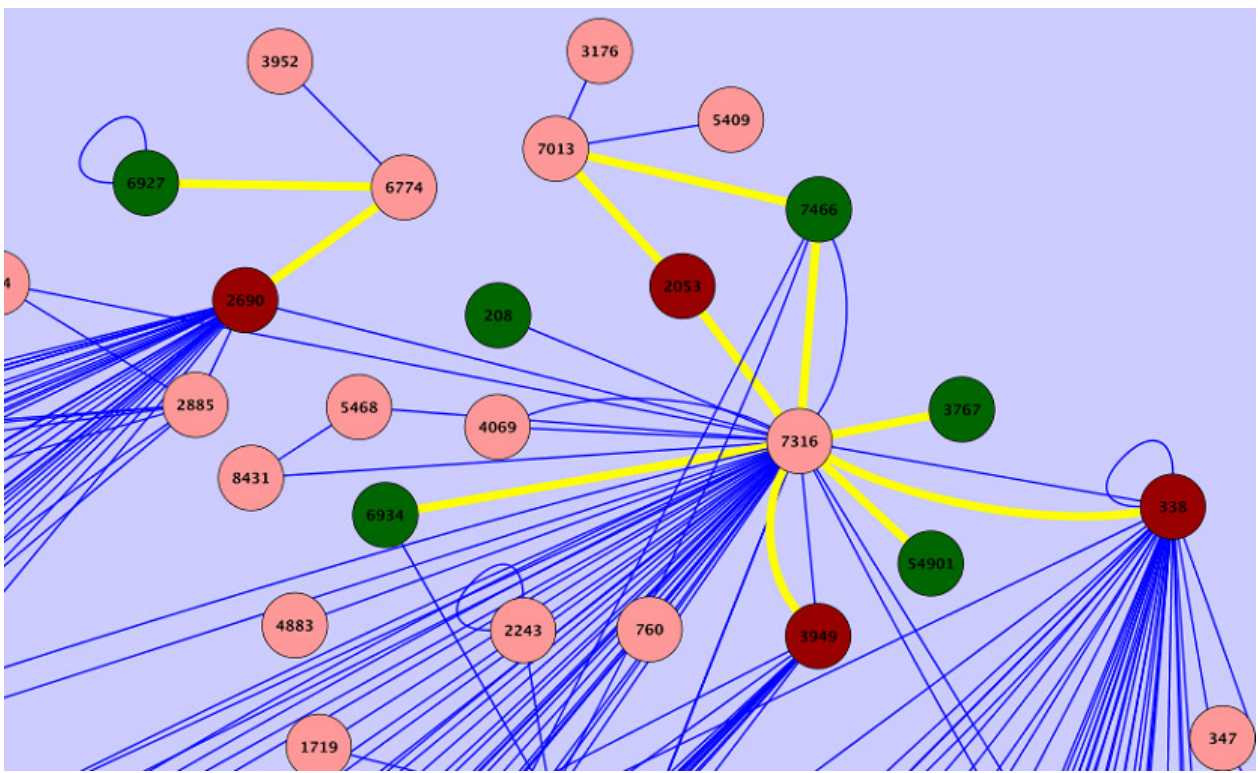

Figure 3: Interaction between genes of Diabetes Mellitus and Hypertension

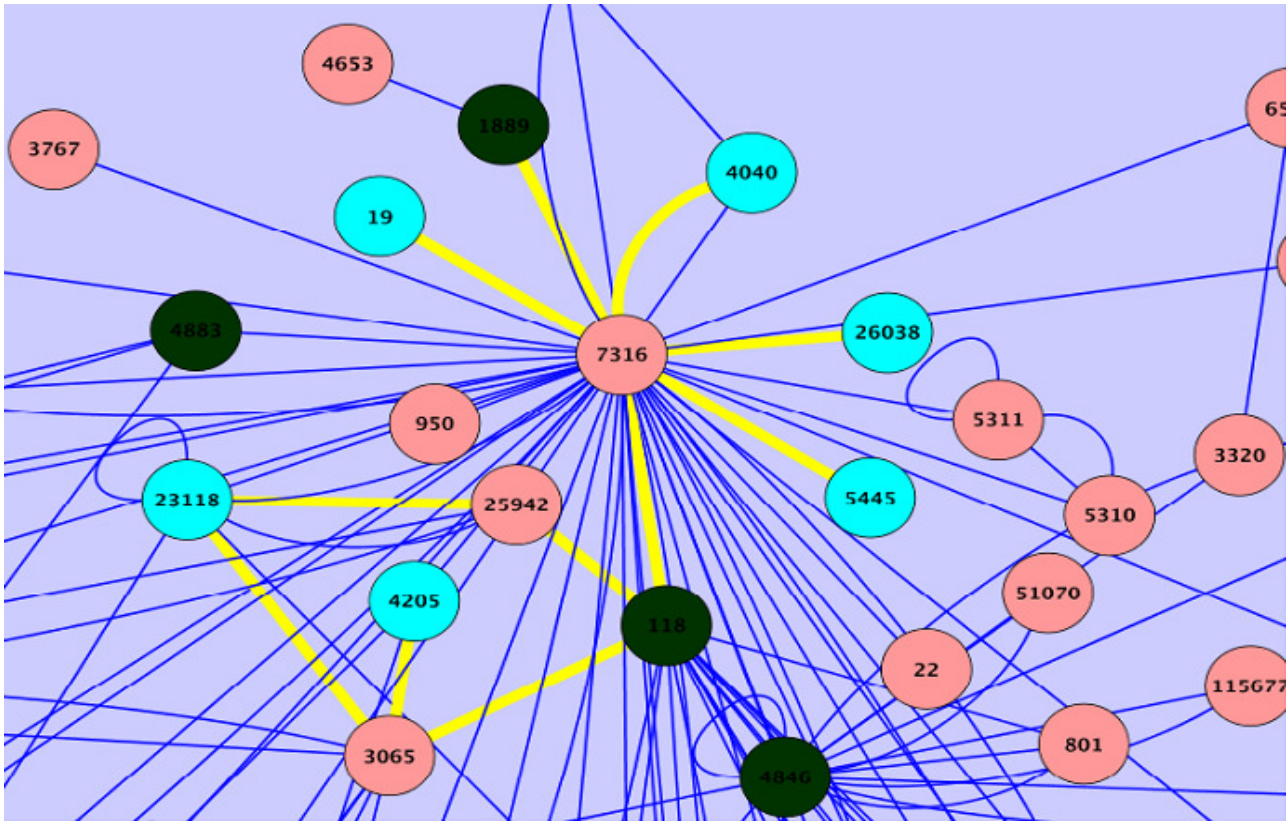

Figure 4: Interaction between genes of Hypertension and Heart Failure

After this final level of validation, following relations were retained as meaningful semantic relation between diseases:

- ***Co-morbid diseases***
  1. Diabetes Mellitus and Hypercholesterolemia
  2. Diabetes Mellitus and Hypertension
  3. Hypercholesterolemia and Hypertension
  4. Obesity and Hypertension

- ***Risk Factors***
  1. Hypertension and Kidney Diseases
  2. Pneumonia and Pulmonary Edema
  3. Obesity and Coronary Heart Disease
  4. Anemia and Left Ventricular Hypertrophy
  5. Hypertension and Heart Failure

**DISCUSSION AND ANALYSIS:**

With meticulous validation against different clinical, bio-medical and molecular sources, some of the derived relations could be established on a strong ground as significant inferences. The acceptability and accuracy of the results obtained through this process was impacted by several constraints in resources. Categorizing diseases through mere section headings alone without involving any precise temporal information and domain expertise is prone to errors. This can be mitigated with suitable usage of precise clinical domain knowledge and inclusion of detailed timeline based data to improve the accuracy. While deducing semantic relations, although co-morbidity could be precisely established, direction of causality for risk factors could not be inferred. Risk factor of a disease is usually not just another disorder, but also multiple aspects like environment, social and economic factors, behavioral traits etc. supplement it. The resources and methods used were not adequate and competent enough to carry out detailed analysis covering so many aspects. There is ample scope for enhancing the accuracy of the core idea through adequate resources like realistic domain expertise and exhaustive knowledge bases.

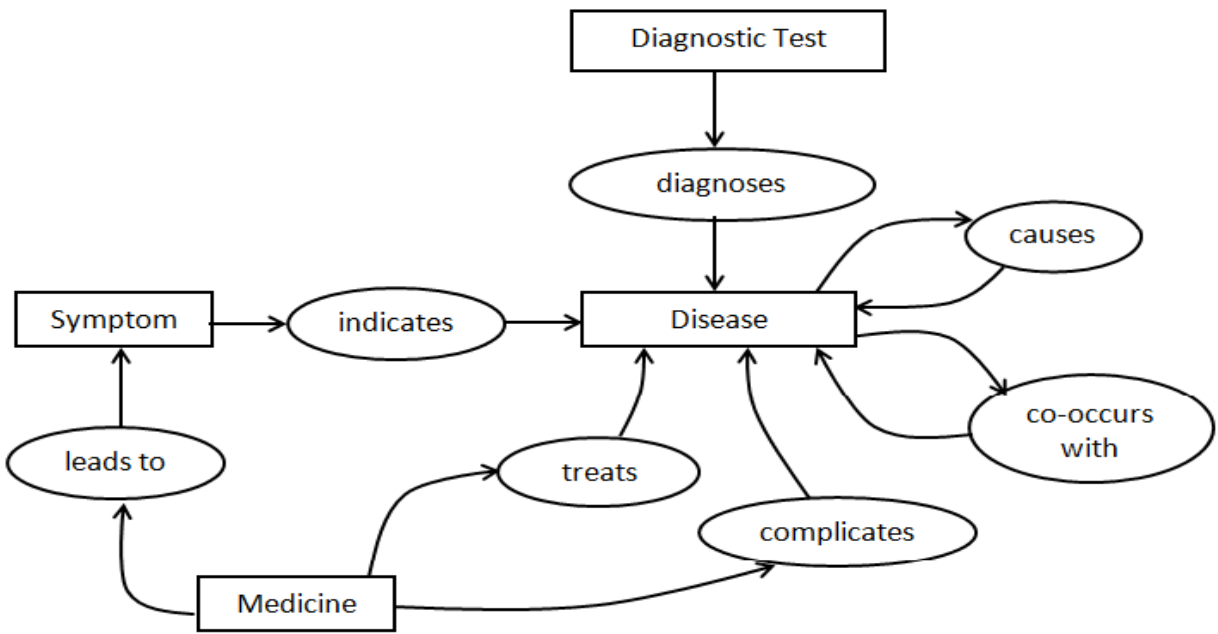

Figure 5: Semantic relations between all types of clinical entities

**CONCLUSION AND FUTURE DIRCTIONS:**

The purpose of this study was to deduce clinical rules from data in such a way that the same can be used in clinical decision making. While deducing clinical knowledge, this method has endeavored to elucidate clinical observations through bio medical knowledge base and genome level networks. Due to constraints arising out of the nature and volume of data and extent of information available in existing knowledge bases, the scope was limited to few well-known diseases. So, clinical rules deduced after appropriate validation were facts which had already been established. But it can be extended to incorporate relatively unknown, complex disorders and uncover their hidden traits to help in clinical assessment of such diseases. Additionally, the core idea can be extended to include all types of clinical entities besides diseases to derive a semantic relation graph as depicted in Figure 5. This will help deducing clinically significant rules to contribute towards better healthcare.

**Acknowledgements:**

Deidentified clinical records used in this research were provided by the i2b2 National Center for Biomedical Computing funded by U54LM008748 and were originally prepared for the Shared Tasks for Challenges in NLP for Clinical Data organized by Dr. Ozlem Uzuner, i2b2 and SUNY.